\documentclass[aps,prd,reprint,nofootinbib,superscriptaddress,amsmath,amssymb]{revtex4-1}

\usepackage[T1]{fontenc}
\usepackage[utf8]{inputenc}
\usepackage{graphicx}
\usepackage{mathrsfs}
\usepackage{amsfonts}
\usepackage{orcidlink}
\usepackage{url}
\hypersetup{hidelinks}
\allowdisplaybreaks[4]

\definecolor{purple}{rgb}{0.58,0.0,0.83}

\definecolor{blue(pigment)}{rgb}{0.2, 0.2, 0.6}
 
\begin{document}

\title{Cosmological Stealth Fields and Non-Equilibrium Thermodynamics}

\author{Gilberto Aguilar-Pérez \orcidlink{0000-0001-6821-4564}}
\email{gilaguilar@uv.mx}
\affiliation{Facultad de F\'{\i}sica, Universidad Veracruzana 91097, Xalapa, Veracruz, M\'exico}

\author{Cuauhtemoc Campuzano \orcidlink{0000-0002-1466-9747}}
\email{ccampuzano@uv.mx}
\affiliation{Facultad de F\'{\i}sica, Universidad Veracruzana 91097, Xalapa, Veracruz, M\'exico}

\author{Víctor H. Cárdenas \orcidlink{0000-0002-9788-3967}}
\email{victor.cardenas@uv.cl}
\affiliation{Instituto de Física y Astronomía, Facultad de Ciencias, Universidad de Valparaíso, Av. Gran Bretaña 1111, Valparaíso, Chile}

\author{Miguel Cruz \orcidlink{0000-0003-3826-1321}}
\email{miguelcruz02@uv.mx}
\affiliation{Facultad de F\'{\i}sica, Universidad Veracruzana 91097, Xalapa, Veracruz, M\'exico}

\author{Joel Saavedra \orcidlink{0000-0002-1430-3008}}
\email{joel.saavedra@pucv.cl}
\affiliation{Instituto de F\'\i sica, Pontificia Universidad Cat\'olica de Valpara\'\i so, Casilla 4950, Valpara\'\i so, Chile}

\date{\today}

\begin{abstract}
We investigate the connection between cosmological stealth scalar fields and non-equilibrium thermodynamics in a spatially flat Friedmann-Lema\^{i}tre-Robertson-Walker (FLRW) background. We consider a non-minimally coupled scalar field whose energy-momentum tensor vanishes identically, allowing the field to evolve on a dissipative cosmological background without producing gravitational backreaction. We show that the stealth condition leads to a generalized Riccati equation for the scalar-field kinematics, where the dissipative pressure acts as a thermodynamic driving term. In terms of the variable $y=\dot{\phi}/(H\phi)$, the system admits two thermodynamic branches: a stable attractor selected by entropy production and an unstable repeller. We also construct the corresponding phase-space structure in the $(y,\omega_{\rm eff})$ plane and identify a near-critical regime associated with $\zeta=1/4$. Finally, we reconstruct the stealth potential and present a bulk viscous realization in which irreversible entropy production drives the universe toward an asymptotic de Sitter state while the stealth field tracks the dissipative background. We further develop the linear scalar perturbations of the model and show that, although the stealth field is gravitationally invisible at the background level, its perturbation couples directly to the dissipative-pressure perturbation. This coupling provides a mechanism for distinguishing dissipative and non-dissipative sources that share the same background expansion history. Our results suggest that stealth fields can be interpreted as dynamically non-trivial thermodynamic trackers of non-equilibrium cosmological evolution.
\end{abstract}


\maketitle

\section{Introduction} 
\noindent 
The realization that the Universe is expanding at an accelerated rate has had a deep impact on contemporary cosmology and remains one of the most significant unresolved questions in gravitational physics. Within the standard cosmological picture, the late-time acceleration is described through the cosmological constant, leading to the successful $\Lambda$CDM model. This scenario provides a phenomenological description of a wide range of observations, including the cosmic microwave background, baryon acoustic oscillations, and type Ia supernovae \cite{planck}. Nevertheless, the physical origin of the dark sector remains unknown, and the cosmological constant problem continues to motivate the exploration of alternative mechanisms that produce accelerated expansion.

A broad class of proposals attempts to address this issue by modifying either the gravitational sector or the effective matter content of the Universe. Scalar fields are among the most widely studied ingredients in this context since they provide a natural framework to model inflation, dynamical dark energy, modified gravity effects, and screening mechanisms. In particular, non-minimally coupled scalar fields appear naturally in many extensions of general relativity and in effective descriptions of gravitational interactions. Their coupling to curvature strongly modifies the cosmological dynamics and produces non-trivial phase-space structures \cite{faraoni2004cosmology}.

An especially interesting possibility arises from stealth scalar configurations. These are non-trivial scalar field profiles whose energy-momentum tensor vanishes on a given spacetime background. Therefore, the scalar field can evolve dynamically while remaining gravitationally invisible. This characteristic turns stealth fields into something much greater than simple mathematical curiosities; they provide a crucial theoretical laboratory for modern cosmology. Exploring the internal dynamics and the thermodynamics of the dark sector typically requires introducing new continuous degrees of freedom. However, standard scalar-tensor approaches inevitably alter the well-constrained geometric background, risking severe tensions with precision cosmological data. Stealth configurations circumvent this limitation. They offer a mathematically rigorous mechanism to track complex, non-equilibrium phenomena such as hidden entropy production or internal energy fluxes while strictly preserving the macroscopic expansion history. Consequently, studying stealth fields allows us to decouple the underlying thermodynamic evolution of these hidden sectors from the gravitational field equations. The present framework may provide a new theoretical avenue to investigate hidden sectors without modifying the gravitational background. Stealth configurations have been investigated in several gravitational contexts, including black holes, lower-dimensional gravity, and cosmological backgrounds \cite{Ayon-Beato:2005yoq,ayonbeato,Ayon-Beato:2013bsa,Ayon-Beato:2015mxf,Ayon-Beato:2024xgp,faraoni,stealthcosmo,Chagoya:2017ojn,campuzano,aguilar2018stealth}. However, their possible connection with irreversible thermodynamics and entropy-producing cosmological fluids remains largely unexplored. On the other hand, the usual perfect-fluid description of the cosmic content is an idealization. Realistic cosmological fluids may experience non-equilibrium processes associated with particle production, bulk viscosity, effective interactions, or departures from local thermodynamic equilibrium. These mechanisms are naturally described within the framework of irreversible thermodynamics, in which dissipative effects are encoded via an effective pressure contribution \cite{PhysRev.58.919, ISRAEL1976310, ISRAEL1979341}. In an expanding Universe, this extra contribution could generate entropy production and may act as a negative pressure source capable of driving accelerated expansion \cite{prigogine,maartens,zim,zimdahl,triginer1996kinetic,lima,waga}. This has motivated extensive studies of viscous cosmology, both in the Eckart formulation and in causal Israel-Stewart-type approaches \cite{Cruz_2017, CRUZ2017159,tamayo,paul2025,Gavassino_2024}.

The aim of the present work is to establish a direct connection between cosmological stealth fields and non-equilibrium thermodynamics. We show that when a non-minimally coupled scalar field satisfies the stealth condition on a dissipative FLRW background, its kinematical evolution is governed by a generalized Riccati equation. In this equation, the dissipative pressure acts as a thermodynamic driving term. As a consequence, the stealth field does not behave as an additional dark energy component and does not modify the gravitational background. Instead, it reacts to the entropy-producing dynamics of the cosmic fluid while preserving its gravitational invisibility.

This framework provides a new interpretation of stealth cosmology. The scalar field behaves as a thermodynamic tracker of the irreversible cosmic evolution. Its evolution is governed by the effective equation of state of the dissipative fluid together with the non-minimal coupling parameter. The resulting system admits thermodynamic branches associated with stable and unstable stealth configurations. We show that the irreversible arrow of time dynamically selects the stable branch, while unstable configurations are repelled. In this sense, entropy production does not merely affect the background expansion; it also determines the physically admissible stealth sector. In addition to the background expansion history, certain modifications to the dark sector can influence the growth of cosmic structures as traced by cosmological perturbations and may help ease existing observational tensions, such as the $\sigma_8$ discrepancy, through changes in the effective gravitational interaction \cite{Gonzalez-Espinoza:2018gyl}. 

A key aspect of the model is the role played by the non-minimal coupling. We identify a critical value of the coupling at which the Riccati structure degenerates into a linear dynamical system. This near-critical regime gives rise to the degeneracy point of the Riccati dynamics in the phase-space structure of the stealth field. The phase portrait reveals how the thermodynamic evolution drives the system toward an asymptotically stable configuration while naturally separating decelerating, accelerating, and effective phantom regimes. At the background level, the effective phantom behavior arises from the dissipative pressure rather than from a scalar field with an explicitly wrong-sign kinetic term. This distinction does not by itself guarantee perturbative stability. We also reconstruct the stealth potential directly from the background stealth condition. We also reconstruct the stealth potential from the stealth condition itself. In this approach, the potential is not specified a priori but is instead fixed by the requirement that the scalar field remains gravitationally invisible. This allows us to interpret the potential as an adaptive structure determined by the background expansion and the thermodynamic state of the cosmic fluid. By projecting the potential onto the phase portrait, we show how the scalar sector evolves across regions with different potential signs and how the potential landscape changes as the non-minimal coupling approaches its critical value.

Finally, we present a concrete bulk viscous realization of the general framework to examine the scope of our formalism. In this example, the dissipative pressure is generated by a simple (constant) bulk viscosity coefficient. The resulting cosmological evolution approaches an asymptotic de Sitter state, while the stealth scalar field is dynamically driven toward its stable thermodynamic branch. This realization illustrates how the general formalism can be implemented in a simple dissipative cosmological model and clarifies the role of the stealth field as a passive but dynamically non-trivial tracker of entropy production.

Throughout this work, we use geometrized units such that $8\pi G=k_{\rm B}=c=1$. Our analysis is performed in the framework of a FLRW spacetime with a scale factor $a(t)$ and a normalized spatial curvature $k=0,\pm1$. Unless otherwise stated, we focus on the spatially flat case, $k=0$, which provides the cosmological background for the dynamical and thermodynamic analysis developed in the following sections.

The paper is organized as follows. In Sec.~\ref{sec:it}, we review the basic ingredients of irreversible thermodynamics in a FLRW cosmological background and introduce the dissipative pressure contribution. In Sec.~\ref{sec:kaniadakis}, we derive the stealth Riccati equation for a non-minimally coupled scalar field and establish its connection with entropy production. We then study the thermodynamic attractors, their stability, and the two-dimensional phase-space structure of the model. We also analyze the near-critical regime associated with the non-minimal coupling and discuss the corresponding stealth potential landscape. In Sec.~\ref{sec:bulk}, we present a bulk viscous realization of the framework and explore its cosmological implications. In Sec. V, we develop the linear scalar perturbation framework for the dissipative stealth configuration.
Finally, in Sec.~\ref{sec:final}, we summarize our main results and discuss possible extensions.

\section{The basics of irreversible thermodynamics}
\label{sec:it}

In the standard cosmological framework, the matter-energy content is typically modeled as a perfect fluid. However, a more realistic description of the cosmological evolution requires the inclusion of non-equilibrium processes. In a FLRW spacetime, dissipative effects can be phenomenologically incorporated by generalizing the energy-momentum tensor $T_{\mu\nu}$ through an effective pressure $P$, such that
\begin{equation}
    T_{\mu\nu} = (\rho + P)u_{\mu}u_{\nu} + P g_{\mu\nu},
\end{equation}
where $\rho$ denotes the energy density, and $u_{\mu}$ is the fluid four-velocity. The total pressure $P$ is decomposed into the local equilibrium pressure $p$, dictated by the equation of state, and a supplementary dissipative contribution $\Pi$, which generates deviations from the local equilibrium pressure, yielding $P = p + \Pi$ \cite{maartens}. When treating the system as a single fluid, the background evolution in the presence of these dissipative effects is determined by the Friedmann equations. For a spatially flat universe, these are given by
\begin{align}
    & 3H^2 = \rho, \label{eq:Friedmann1} \\
    & 2\dot{H} + 3H^2 = -P = -p-\Pi, \label{eq:Friedmann2}
\end{align}
where $H = \dot{a}/a$ is the Hubble parameter, and the dot stands for derivatives with respect to cosmic time. In general, the dissipative term satisfies $\Pi = \Pi(t)$. The equation \eqref{eq:Friedmann2} clearly shows that the dissipative contribution $\Pi$ has a direct impact on cosmic acceleration. If $\Pi$ is sufficiently negative, such that $\rho + 3(p + \Pi) < 0$, the dissipative mechanism can drive an accelerated expansion phase without the need for extra contributions. Furthermore, the dissipative term modifies the conservation equation of equilibrium energy in the following way
\begin{equation}
    \dot{\rho} + 3H(\rho + p + \Pi) = 0.\label{eq:noncons}
\end{equation}
From the perspective of irreversible thermodynamics, the presence of $\Pi$ is intrinsically linked to entropy production. Entropy-generating processes can be modeled using dissipative fluid descriptions. Among the standard dissipative mechanisms—heat conduction, shear viscosity, and bulk viscosity—bulk viscosity is the preferred option in cosmology, as it is the only one consistent with the symmetry conditions of homogeneous and isotropic FLRW universes \cite{Zimdahl_1997}. Another intriguing way to generate entropy, appearing as an additional pressure term, is through particle creation from spacetime itself, a mechanism that has been widely investigated in the literature for late times cosmic evolution; see, for instance, the Refs. given in \cite{waga, Cardenas_2024, schiavone2026}. The Gibbs relation
\begin{equation}
    TdS = dU+pdV, \label{eq:gibss}
\end{equation}
may be expressed in terms of the dissipative term as follows
\begin{equation}
    \frac{T}{V}\frac{dS}{dt} =-3H\Pi, \label{eq:entropy}
\end{equation}
where Eq. (\ref{eq:noncons}) was considered. In the above expressions, $U$ is the internal energy defined as $U=\rho V$, $V = V_{0}a^{3}(t)$ is the comoving volume, and $T$ is the temperature of the cosmic fluid. This outcome indicates that the system’s departure from adiabaticity—and thus its entropy production—is fully determined by the dissipative quantity $\Pi$. Because we do not introduce any extra fluid components with independent thermodynamic degrees of freedom, all entropy production is encoded solely in $\Pi$. From Eq. \eqref{eq:entropy}, it follows that to ensure positive entropy production, we must impose $\Pi < 0$, because in an expanding universe we have $H, V > 0$, and for physical consistency, the fluid temperature is assumed to be positive. Observe that, once the form of $\Pi$ has been determined, a number of physically relevant cases can be analyzed.

\section{Stealth cosmology}
\label{sec:kaniadakis}
The energy–momentum tensor for a scalar field that is non-minimally coupled to gravity via the term $-\frac{1}{2}\zeta R\phi^2$ includes contributions proportional to the Einstein tensor $G_{\mu\nu}$. Following \cite{campuzano}, in a spatially flat FLRW universe, enforcing the null energy-momentum tensor, $T_{00}^{(S)} = 0$, the time component leaves
\begin{equation}
    3H^2 = -6Hx - \frac{1}{2\zeta}x^2 - \frac{V}{\zeta\phi^2}, \label{eq:stealth_00}
\end{equation}
where $x:=\dot{\phi}/\phi$, and the spatial terms, $T_{ij}^{(S)} = 0$, 
\begin{equation}
    2\dot{H} + 3H^2 = -\frac{V}{\zeta\phi^{2}} + \frac{1}{2\zeta}x^{2} - 4Hx - 4x^{2} - 2\dot{x}. \label{eq:stealth_ij}
\end{equation}
From the background Friedmann equations, the spatial acceleration is dictated by the effective thermodynamic pressure \eqref{eq:Friedmann2}, $2\dot{H} + 3H^2 = -p - \Pi$. Substituting this geometric background into the left side of Eq. \eqref{eq:stealth_ij} yields:
\begin{equation}
     -p - \Pi = -\frac{V}{\zeta\phi^2} + \frac{1}{2\zeta}x^2 - 4Hx - 4x^2 - 2\dot{x}.  \label{eq:spatial_sub}
\end{equation}
To close the system and eliminate the explicit dependence on the potential $V(\phi)$, we solve for $V/\zeta\phi^2$ from \eqref{eq:stealth_00}, and upon inserting it into \eqref{eq:spatial_sub}, one gets
\begin{eqnarray}
    -p - \Pi &=& \left( 3H^2 + 6Hx + \frac{1}{2\zeta}x^2 \right) + \frac{1}{2\zeta}x^2 \nonumber \\
    && - 4Hx - 4x^2 - 2\dot{x},
\end{eqnarray}
which in turn results as
\begin{equation}
    -p - \Pi = 3H^2 + 2Hx + \left(\frac{1}{\zeta} - 4\right)x^2 - 2\dot{x}.
\end{equation}
Finally,  we obtain a Riccati equation for the stealth field
\begin{equation}
    \dot{x} + \left(2 - \frac{1}{2\zeta}\right)x^2 - Hx - \frac{3}{2}H^2 -\frac{1}{2}p= \frac{1}{2}\Pi. \label{eq:generalized_riccati}
\end{equation}
This latter equation forms the foundation of our unified framework. In mathematical terms, it is a nonlinear Riccati differential equation driven by the thermodynamic state of the universe. From a physical perspective, it recasts the stealth kinematics in terms of the background’s irreversible thermodynamic evolution. We stress that, at this level, the field does not generate the thermodynamics: the combination $2\dot{H}+3H^{2}=-P$ shows that only the total pressure enters, so the same equation would follow from any source producing the same expansion history. The physically distinctive content emerges through the thermodynamic selection of the stable branch, the constraint on $\zeta$ derived below, and the perturbative analysis as we will show. The dissipative pressure $\Pi$ acts as an external driving force or source term. When the universe expands adiabatically ($\Pi = 0$), the kinematics of the scalar field are solely governed by the geometric expansion $H$ and the equilibrium pressure $p$. However, the presence of dissipation disrupts this behavior. As discussed in the previous section, the second law of thermodynamics requires $\Pi < 0$ to ensure positive entropy production during expansion. Consequently, the term $\Pi/2$ on the right-hand side is strictly negative, so dissipation enters as a negative source term in the stealth kinematics.

\subsection{Stealth kinematics and entropy production}
Now, Eq. (\ref{eq:generalized_riccati}) relates the non-equilibrium thermodynamics of a generic dissipative fluid to the dynamics of the stealth. In order to integrate the Riccati equation, we propose 
a barotropic effective equation of state for the cosmic fluid, $\omega_{\text{eff}}(t) = (p + \Pi)/\rho$. Therefore, the total pressure takes the form
\begin{equation}
    \frac{1}{2}(p + \Pi) = \frac{3}{2}H^2 \omega_{\text{eff}}. \label{eq:effective_pressure_term}
\end{equation}
Additionally, the temporal evolution of the Hubble parameter is determined by the geometric acceleration equation $2\dot{H} + 3H^2 = -(p + \Pi)$. Substituting Eq. \eqref{eq:effective_pressure_term} yields
\begin{equation}
    2\dot{H} = -3H^2 - 3H^2\omega_{\text{eff}} \implies \dot{H} = -\frac{3}{2}H^2(1 + \omega_{\text{eff}}). \label{eq:h_dot_general}
\end{equation}
By using a dimensionless variable, $y = x/H$, Eq. (\ref{eq:generalized_riccati}) takes the form
\begin{equation}
    y' + \gamma y^2 - \left(\frac{5}{2} + \frac{3}{2}\omega_{\text{eff}}\right)y - \frac{3}{2}\left(1 + \omega_{\text{eff}}\right)=0. \label{eq:riccati_general}
\end{equation}
Here, we employed the results obtained above and introduced the effective coupling constant $\gamma = 2 - 1/(2\zeta)$. Dividing by $H^2$, we can establish the relation $\dot{y}/H=(dy/dt)(dt/da) a=dy/d(\ln a)$. We therefore define the derivative $y' \equiv dy/d (\ln a)$. As commented before, the net entropy production rate is driven exclusively by the presence of the non-equilibrium pressure $\Pi$
\begin{equation}
    T\dot{S} = -3HV\Pi \implies dS = -\frac{3V\Pi}{T} \frac{da}{a} = -\frac{3V\Pi}{T} d(\ln a). \label{eq:entropy2}
\end{equation}
Using the results obtained above, Eq. \eqref{eq:riccati_general} yields the following analytical relation
\begin{equation}
    \frac{dy}{dS} = \frac{T}{3V\Pi} \left[ \gamma y^2 - \left(\frac{5}{2} + \frac{3}{2}\omega_{\text{eff}}\right)y - \frac{3}{2}\left(1 + \omega_{\text{eff}}\right)\right]. \label{eq:universal_entropic_riccati}
\end{equation}
The connection between the scalar field and the entropy of the universe is strictly given by the term $dy/dS$ in Eq. \eqref{eq:universal_entropic_riccati}. Physically, this implies that the field cannot instantly adjust to sudden changes in the cosmic dissipation rate $\Pi$. Instead, the field exhibits a ``dynamical inertia'' and evolves in the direction determined by the thermodynamic arrow of time. This finding is consistent with the conclusions of \cite{quevedo2024timeentropygeometricproof}, which demonstrate that the arrow of time is determined by the direction in which entropy increases. The non-minimal coupling constant $\zeta$ regulates this thermodynamic ``stiffness'', dictating how rapidly the field's kinematics can respond to the dissipative forces driving the universe. Within the assumptions of spatial flatness, a single effective fluid, local equilibrium variables, and the background stealth condition, Eq. (17) retains the same form independently of the specific mechanism generating the dissipative pressure. The underlying transport physics nevertheless determines the evolution of \(\Pi\), \(T\), and \(\omega_{\mathrm{eff}}\). In the adiabatic limit $\Pi \to 0$, entropy is no longer a suitable time variable for the stealth dynamics. The evolution with respect to $N=\ln a$ may still be well defined, but it is no longer driven by entropy production.

The potential \(V(\phi)\) is not specified a priori. Instead, it is reconstructed from the background stealth condition so that the scalar configuration remains gravitationally invisible throughout the cosmological evolution. Using the previous results, we obtain
    \begin{equation}
        V(\phi) = -\zeta\phi^2 H^2 \left( 6y + \frac{1}{2\zeta}y^2 + 3 \right).
    \end{equation}

\subsection{Instantaneous branches and asymptotic states}

For a constant value of \(\omega_{\mathrm{eff}}\), Eq. (17) defines an autonomous one-dimensional dynamical system for the kinematic variable \(y(S)\). More generally, when \(\omega_{\mathrm{eff}}\) evolves with time, the roots obtained by imposing \(dy/dS=0\) define instantaneous equilibrium branches, or nullclines, rather than fixed points of the complete cosmological system. Genuine asymptotic fixed points arise when \(\omega_{\mathrm{eff}}\) itself approaches a constant value. Under either the constant-\(\omega_{\mathrm{eff}}\) assumption or at such an asymptotic state, the equilibrium condition is
\begin{equation}
    \frac{dy}{dS} = 0.
\end{equation}
The above condition, assuming a non-zero thermodynamic driving factor ($T/(3V\Pi) \neq 0$), is satisfied when the term in the square brackets of \eqref{eq:universal_entropic_riccati} vanishes
\begin{equation}
    \gamma y_*^2 - \left(\frac{5}{2} + \frac{3}{2}\omega_{\text{eff}}\right)y_* - \frac{3}{2}\left(1 + \omega_{\text{eff}}\right) = 0, \label{eq:fixed_points_condition}
\end{equation}
at a state denoted by \(y_*\). Solving this quadratic constraint yields two instantaneous equilibrium branches for the stealth kinematics. When \(\omega_{\mathrm{eff}}\) is constant, or when it approaches an asymptotic constant value, these branches reduce to the fixed points
\begin{equation}
    y_{*}^{\pm} = \frac{\left(\frac{5}{2} + \frac{3}{2}\omega_{\text{eff}}\right) \pm \sqrt{\left(\frac{5}{2} + \frac{3}{2}\omega_{\text{eff}}\right)^2 + 6\gamma\left(1 + \omega_{\text{eff}}\right)}}{2\gamma}. \label{eq:attractor_roots}
\end{equation}
Equation \eqref{eq:attractor_roots} shows that the location of the instantaneous equilibrium branches is determined by the non-minimal coupling parameter \(\gamma\) and the current value of the effective equation-of-state parameter \(\omega_{\mathrm{eff}}\). The asymptotic fate of the stealth field additionally depends on the evolution law governing \(\omega_{\mathrm{eff}}\).
For the instantaneous equilibrium branches in Eq. \eqref{eq:attractor_roots} to be real, their discriminant must be strictly positive
    \begin{equation}
        \Delta = \left(\frac{5}{2} + \frac{3}{2}\omega_{\text{eff}}\right)^2 + 6\gamma\left(1 + \omega_{\text{eff}}\right) \geq 0. \label{eq:discriminant_condition}
    \end{equation}
This condition constrains the combinations of \(\zeta\) and \(\omega_{\mathrm{eff}}\) for which real instantaneous equilibrium branches exist. If the inequality is violated, Eq. \eqref{eq:fixed_points_condition} has no real roots at that value of \(\omega_{\mathrm{eff}}\). This does not imply that the time-dependent stealth solution ceases to exist; rather, it means that the kinematic variable \(y\) has no instantaneous stationary value in that region of parameter space.
\subsubsection{Stability Analysis and the Thermodynamic Selection}
When \(\omega_{\mathrm{eff}}\) is constant, the quadratic constraint \eqref{eq:fixed_points_condition} yields two fixed points, \(y_{*+}\) and \(y_{*-}\), whose stability can be determined directly. If \(\omega_{\mathrm{eff}}\) varies, the same calculation determines the local stability of the corresponding instantaneous branches in the \(y\) direction, but not the stability of the complete cosmological system. For constant \(\omega_{\mathrm{eff}}\), a fixed point \(y_*\) is a stable attractor if a small kinematic perturbation \(\delta y\) decays with increasing entropy, which requires
\begin{equation}
    \left. \frac{d}{dy} \left( \frac{dy}{dS} \right) \right|_{y_*} < 0.
\end{equation}
Let $f(y)$ denote the right-hand side of our Riccati Equation \eqref{eq:universal_entropic_riccati}. Differentiating $f(y)$ with respect to the variable $y$ yields
\begin{equation}
    f'(y) = \frac{T}{3V\Pi} \left[ 2\gamma y - \left(\frac{5}{2} + \frac{3}{2}\omega_{\text{eff}}\right) \right].
\end{equation}
When this derivative is evaluated at the roots specified in Eq. \eqref{eq:attractor_roots}, the bracketed term reduces precisely to the square root of the discriminant defined above
\begin{equation}
    2\gamma y_{*}^{\pm} - \left(\frac{5}{2} + \frac{3}{2}\omega_{\text{eff}}\right) = \pm \sqrt{\Delta}.
\end{equation}
Thus, the eigenvalues governing the stability of the two steady states are given as follows
\begin{equation}
    \lambda_{\pm} = f'(y_{*}^{\pm}) = \pm \frac{T}{3V\Pi}\sqrt{\Delta}.
\end{equation}
As established in the previous section, the second law of thermodynamics for an expanding universe ($dS > 0$, $H > 0$) strictly requires the dissipative pressure to be negative, $\Pi < 0$. Since the absolute temperature $T$ and the comoving volume $V$ are strictly positive quantities, the global factor $T/(3V\Pi)$ is negative.
Since the square root $\sqrt{\Delta}$ is positive by definition (for real roots), the signs of the eigenvalues are completely determined
\begin{align}
    \lambda_{+} &= \frac{T}{3V\Pi} \left( + \sqrt{\Delta} \right) < 0 \implies \textbf{Stable attractor}, \\
    \lambda_{-} &= \frac{T}{3V\Pi} \left( - \sqrt{\Delta} \right) > 0 \implies \textbf{Unstable repeller}.
\end{align}
For constant \(\omega_{\mathrm{eff}}\) and positive entropy production, this stability analysis shows that \(y_{*+}\) is locally attractive in the \(y\) direction, whereas \(y_{*-}\) is repulsive. When \(\omega_{\mathrm{eff}}\) evolves, the same conclusion applies locally to the instantaneous branches, provided that they remain real and vary sufficiently slowly compared with the relaxation of \(y\). The negative dissipative pressure fixes the orientation between entropic time and cosmological time, thereby determining which of the two branches is attractive. It should not, however, be interpreted as a conventional friction coefficient in the Riccati equation, for $p=0$ we have 
\begin{equation}
    \omega_{\text{eff}} = \frac{\Pi}{\rho}.
\end{equation}
Since the second law dictates $\Pi < 0$, the effective equation of state is strictly negative. Substituting this into our thermodynamic bound (\ref{eq:discriminant_condition}) yields a direct constraint between the scalar field coupling $\zeta$ and the macroscopic dissipation $\Pi$
\begin{equation}
    2 - \frac{1}{2\zeta} \geq -\frac{\left(\frac{5}{2} + \frac{3}{2}\frac{\Pi}{\rho}\right)^2}{6\left(1 + \frac{\Pi}{\rho}\right)}.\label{eq:bound}
\end{equation}
This inequality reveals how the coupling parameter and the dissipative effects come into play. In the de Sitter limit $\Pi \to -\rho$ ($\omega_{\text{eff}} \to -1$), the discriminant reduces to $\Delta = 1 > 0$ for any value of $\zeta$, so the stealth condition can always be satisfied and no bound on the coupling arises in this limit; in particular, the conformal value $\zeta = 1/6$ is not excluded. A nontrivial constraint appears instead in the phantom regime $\omega_{\text{eff}} < -1$, where $1+\omega_{\text{eff}} < 0$ so that the inequality \eqref{eq:bound} reverses and becomes an upper bound on the coupling. Finally, the thermodynamic bound \eqref{eq:bound} provides a simple mechanism for accommodating an effective phantom regime within the background stealth dynamics. In conventional scalar-field models, an equation-of-state parameter \(\omega_{\mathrm{eff}}<-1\) is often associated with a wrong-sign kinetic term. In the present construction, by contrast, the phantom behavior is generated by the macroscopic dissipative pressure, \(\Pi<-\rho\), rather than by assigning a negative kinetic term to the stealth scalar field. The field consequently shifts its kinematic equilibrium in response to the entropy-producing background. For instance, when \(\omega_{\mathrm{eff}}=-1.1\), the discriminant remains non-negative for \(\zeta\lesssim0.63\), ensuring the existence of real stealth branches. This background result, however, does not by itself establish the absence of ghost or gradient instabilities, which requires a separate perturbative analysis. In contrast, our thermodynamic framework allows the universe to enter a phantom regime driven purely by extreme macroscopic dissipation ($\Pi <-\rho$) within the cosmic fluid. The stealth scalar field requires no modifications; its Lagrangian remains theoretically well-behaved. The field simply shifts its kinematic equilibrium to absorb the severe entropy production. As dictated by our discriminant bound, an effective phantom state (for instance, $\omega_{\text{eff}}=-1.1$) remains physically viable as long as the coupling is not too large, $\zeta \lesssim 0.63$, so that the roots stay real without invoking negative kinetic energies.

\subsubsection{Two-dimensional phase-space structure}

From the perspective of irreversible thermodynamics, there is no fundamental reason for the effective equation of state $\omega_{\rm eff}$ to remain constant during cosmological evolution. Since the dissipative pressure $\Pi$ may evolve dynamically as entropy is produced, the effective thermodynamic state of the cosmic fluid can naturally become time-dependent. This motivates the construction of a two-dimensional autonomous dynamical system in which both the stealth kinematic variable and the thermodynamic state evolve simultaneously.

To this end, we promote the effective equation of state to an independent dynamical variable and define the phase-space coordinates
\begin{equation}
    \left(y,\omega_{\rm eff}\right),
\end{equation}
where, as before,
\begin{equation}
    y \equiv \frac{\dot{\phi}}{H\phi},
    \qquad
    N \equiv \ln a.
\end{equation}
Under the background assumptions specified above, the stealth kinematic equation can be rewritten as
\begin{equation}
    y'
    =
    -\gamma y^2
    +
    \left(
    \frac{5}{2}
    +
    \frac{3}{2}\omega_{\rm eff}
    \right)y
    +
    \frac{3}{2}\left(1+\omega_{\rm eff}\right),
    \label{eq:phase_y}
\end{equation}
where the prime denotes derivatives with respect to $N$ and
\begin{equation}
    \gamma = 2 - \frac{1}{2\zeta}.
\end{equation}

To close the system without specifying any particular dissipative mechanism, we introduce a general evolution law for the effective thermodynamic parameter,
\begin{equation}
    \omega_{\rm eff}' = \mathcal{F}(\omega_{\rm eff}),
    \label{eq:phase_omega}
\end{equation}
where the function $\mathcal{F}$ encodes the underlying non-equilibrium transport physics. Different realizations of dissipation, such as bulk viscosity, particle production, or causal transport theories, correspond to different choices of $\mathcal{F}$. Therefore, Eqs. \eqref{eq:phase_y} and \eqref{eq:phase_omega} define a partially model-independent phase-space formulation: the evolution equation for the stealth variable has a fixed functional form under the assumptions of the present framework, whereas the thermodynamic evolution remains model dependent through \(F(\omega_{\mathrm{eff}})\):

\begin{equation}
    \begin{aligned}
    y'
    &=
    -\gamma y^2
    +
    \left(
    \frac{5}{2}
    +
    \frac{3}{2}\omega_{\rm eff}
    \right)y
    +
    \frac{3}{2}\left(1+\omega_{\rm eff}\right),
    \\
    \omega_{\rm eff}'
    &=
    \mathcal{F}(\omega_{\rm eff}).
    \end{aligned}
\end{equation}

The critical points of the system satisfy
\begin{equation}
    y' = 0,
    \qquad
    \omega_{\rm eff}' = 0.
\end{equation}
Consequently, the equilibrium thermodynamic states are determined by the roots of
\begin{equation}
    \mathcal{F}(\omega_*) = 0,
\end{equation}
while the corresponding stealth equilibria are given by
\begin{equation}
    y_*^{\pm} =
    \frac{
    \left(
    \frac{5}{2}
    +
    \frac{3}{2}\omega_*
    \right)
    \pm
    \sqrt{
    \left(
    \frac{5}{2}
    +
    \frac{3}{2}\omega_*
    \right)^2
      +
    6\gamma
    \left(
    1
    +
    \omega_*
    \right)
    }
    }
    {2\gamma}.
\end{equation}
Therefore, each asymptotic thermodynamic state $\omega_*$ generates two equilibrium points in the phase plane,
\begin{equation}
    (y_*^+,\omega_*),
    \qquad
    (y_*^-,\omega_*).
\end{equation}

The local stability of the system is determined through the Jacobian matrix
\begin{equation}
    J =
    \begin{pmatrix}
    -2\gamma y
    +
    \frac{5}{2}
    +
    \frac{3}{2}\omega_{\rm eff}
    &
    \frac{3}{2}(y+1)
    \\
    0
    &
    \mathcal{F}'(\omega_{\rm eff})
    \end{pmatrix}.
\end{equation}
Evaluating the Jacobian at the critical points yields the eigenvalues
\begin{equation}
    \lambda_1^{\pm}
    =
    -2\gamma y_*^{\pm}
    +
    \frac{5}{2}
    +
    \frac{3}{2}\omega_*,
\end{equation}
and
\begin{equation}
    \lambda_2
    =
    \mathcal{F}'(\omega_*).
\end{equation}
Using the explicit form of the roots, the first eigenvalue simplifies to
\begin{equation}
    \lambda_1^{\pm}
    =
    \mp\sqrt{
    \left(
    \frac{5}{2}
    +
    \frac{3}{2}\omega_*
    \right)^2
    +
    6\gamma
    \left(
    1
    +
    \omega_*
    \right)
    }.
\end{equation}
Thus, for real and hyperbolic critical points, the branch \(y_{*+}\) is locally stable along the stealth direction, whereas \(y_{*-}\) is unstable in that direction. The stability of the complete cosmological equilibrium also depends on the thermodynamic eigenvalue \(F'(\omega_*)\). In particular, the point \((y_{*+},\omega_*)\) is a local sink when $\mathcal{F}'(\omega_*) < 0$.

This two-dimensional formulation separates the universal evolution of the stealth variable from the model-dependent thermodynamic evolution encoded in $\mathcal{F}(\omega_{\mathrm{eff}})$. At a thermodynamic fixed point satisfying $\mathcal{F}(\omega_*) = 0$, the branch \(y_{*-}\) is repulsive in the stealth direction, whereas \(y_{*+}\) is attractive in that direction. The latter becomes a local sink of the complete system only when the thermodynamic direction is also stable, namely when $\mathcal{F}'(\omega_*) < 0$.

\subsubsection{Near-critical phase portrait and degeneracy point of the Riccati dynamics}

\begin{center}
\begin{figure*}[htbp!]
    \includegraphics[scale=0.35]{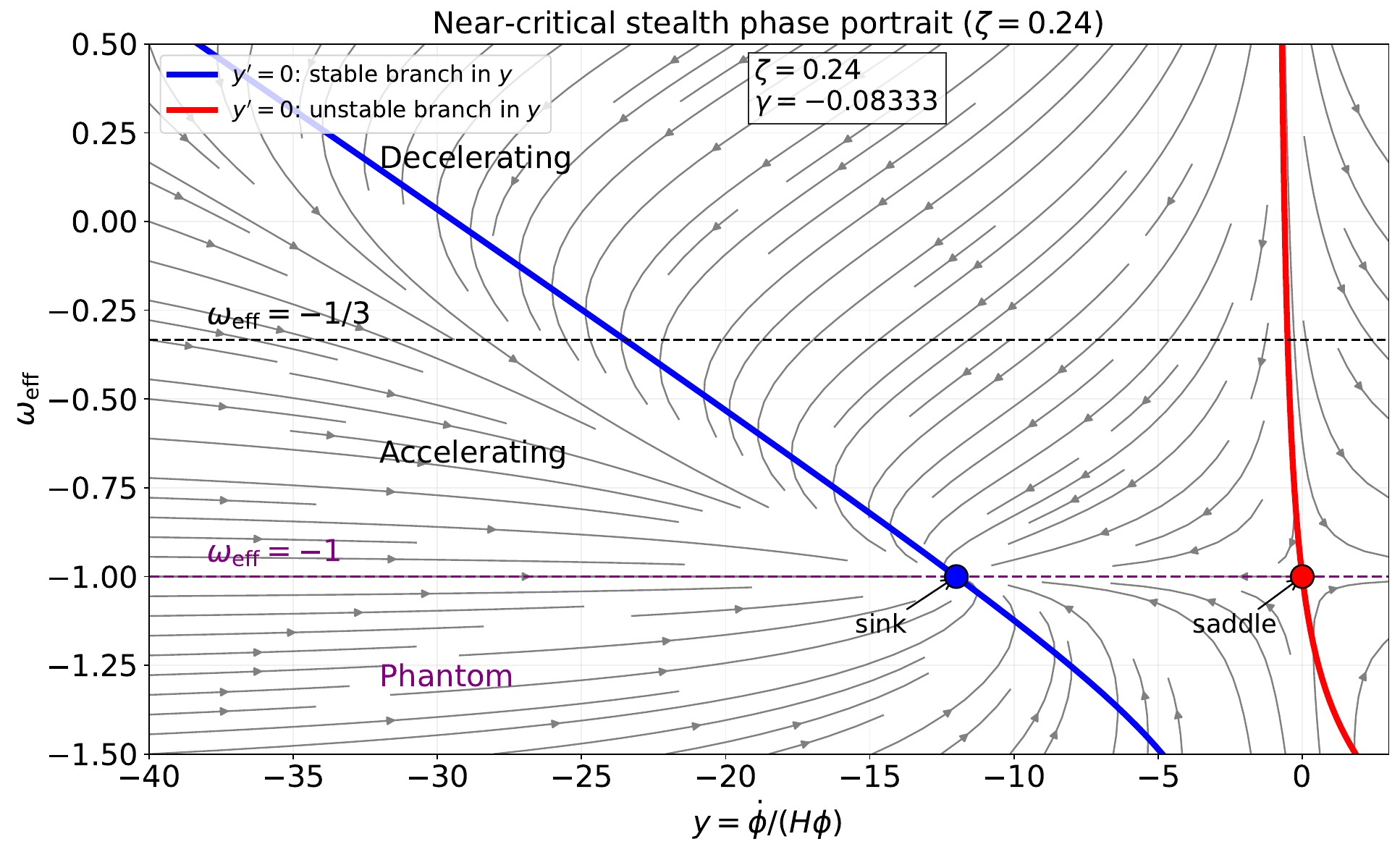}
    \caption{Two-dimensional phase portrait in the $(y,\omega_{\rm eff})$ plane for the near-critical regime $\zeta=0.24$. The blue curve corresponds to the stable stealth branch $y_*^+(\omega_{\rm eff})$, while the red curve represents the unstable branch $y_*^-(\omega_{\rm eff})$. The horizontal dashed lines indicate the transition to accelerated expansion at $\omega_{\rm eff}=-1/3$ and the de Sitter invariant submanifold at $\omega_{\rm eff}=-1$. The flow demonstrates how the irreversible thermodynamic evolution dynamically drives the system toward the stable stealth configuration.}
    \label{fig:phase_portrait}
\end{figure*}
\end{center}

A particularly interesting dynamical regime emerges when the non-minimal coupling parameter approaches the critical value
\begin{equation}
    \zeta_c=\frac{1}{4}.
\end{equation}
At this point, the effective coupling coefficient $\gamma$ vanishes, and the nonlinear Riccati structure degenerates into a linear dynamical equation. Consequently, the structure of the phase space changes qualitatively, revealing the existence of a 
degeneracy point of the Riccati dynamics in the stealth dynamics.
To illustrate this behavior, we consider a near-critical configuration with
\begin{equation}
    \zeta = 0.24,
\end{equation}
together with the phenomenological relaxation law
\begin{equation}
    \omega_{\rm eff}'=
    -\alpha(\omega_{\rm eff}+1),\label{eq:relax}
\end{equation}
where $\alpha>0$. This choice drives the effective thermodynamic state asymptotically toward the de Sitter invariant submanifold $\omega_{\rm eff}=-1$.
For $\alpha=1$, the resulting phase portrait in the $(y,\omega_{\rm eff})$ plane exhibits several remarkable features, as can be seen in Figure \ref{fig:phase_portrait}. First, the trajectories are globally attracted towards the de Sitter line, confirming that the irreversible thermodynamic evolution dynamically selects an asymptotic accelerated state. However, close to the critical coupling $\zeta_c=1/4$, the stealth sector develops a highly asymmetric phase-space structure.
\onecolumngrid
\twocolumngrid
For \(\alpha=1\), the trajectories displayed in the \((y,\omega_{\mathrm{eff}})\) phase portrait approach the invariant de Sitter line \(\omega_{\mathrm{eff}}=-1\), consistently with the relaxation law introduced in Eq. (\ref{eq:relax}).
Within the region and initial conditions represented in Fig. (\ref{fig:phase_portrait}), the flow approaches the local sink at \(y=1/\gamma\). For \(\zeta=0.24\), corresponding to \(\gamma\simeq-0.0833\), this de Sitter fixed point is located at \(y\simeq-12\). Near the critical coupling $\zeta_c=1/4$ the two branches behave oppositely: since both roots scale as $1/\gamma$, the limit $\gamma\to0$ is singular. The unstable branch $y_*^-$ develops a cancellation in its numerator and remains finite, $y_*^-\to-\,3(1+\omega_{\rm eff})/(5+3\omega_{\rm eff})$ (near $y\simeq0$), whereas the stable branch $y_*^+$ is displaced to large $|y|$.

This behavior can be understood analytically from the critical solutions $y_*^{\pm}$. Since both solutions scale as $1/\gamma$, the limit $\gamma\to0$ becomes singular. One branch remains finite due to a cancelation in the numerator, while the other diverges. Therefore, the near-critical regime corresponds to a structural degeneracy point of the Riccati dynamics in the stealth phase space.

From a physical viewpoint, near the critical coupling the thermodynamic evolution retains a single finite equilibrium, while the stable attractor recedes to arbitrarily large $|y|$,
forcing the scalar field to roll increasingly fast in order to keep
tracking the entropy-producing background. In this sense, the
irreversible entropy production not only determines the direction of
cosmological evolution but also governs the global structure of the
stealth phase space.

Another remarkable aspect of the phase portrait is the natural separation between different cosmological regimes. The line $\omega_{\rm eff}=-\frac13$ divides decelerating and accelerating expansion, while the region $\omega_{\rm eff}<-1$ corresponds to an effective phantom regime entirely induced by dissipative entropy production. Importantly, the effective phantom regime is generated by the dissipative pressure rather than by introducing a scalar field with an explicitly wrong-sign kinetic term. The regularity of the background stealth trajectory does not, however, establish the perturbative stability or causal consistency of the complete system.

Finally, the phase portrait reveals that the critical value $\zeta=1/4$ plays the role of a dynamical transition point, separating qualitatively different stealth cosmologies. For $\zeta>1/4$, the system exhibits two nonlinear Riccati branches, 
while in the vicinity of $\zeta\simeq1/4$ the stable branch is driven off to infinity, so that the dynamics is effectively governed by a single attractor.

\subsubsection{Phase portrait and stealth potential landscape}
\begin{center}
\begin{figure*}[htbp!]
    \includegraphics[scale=0.4]{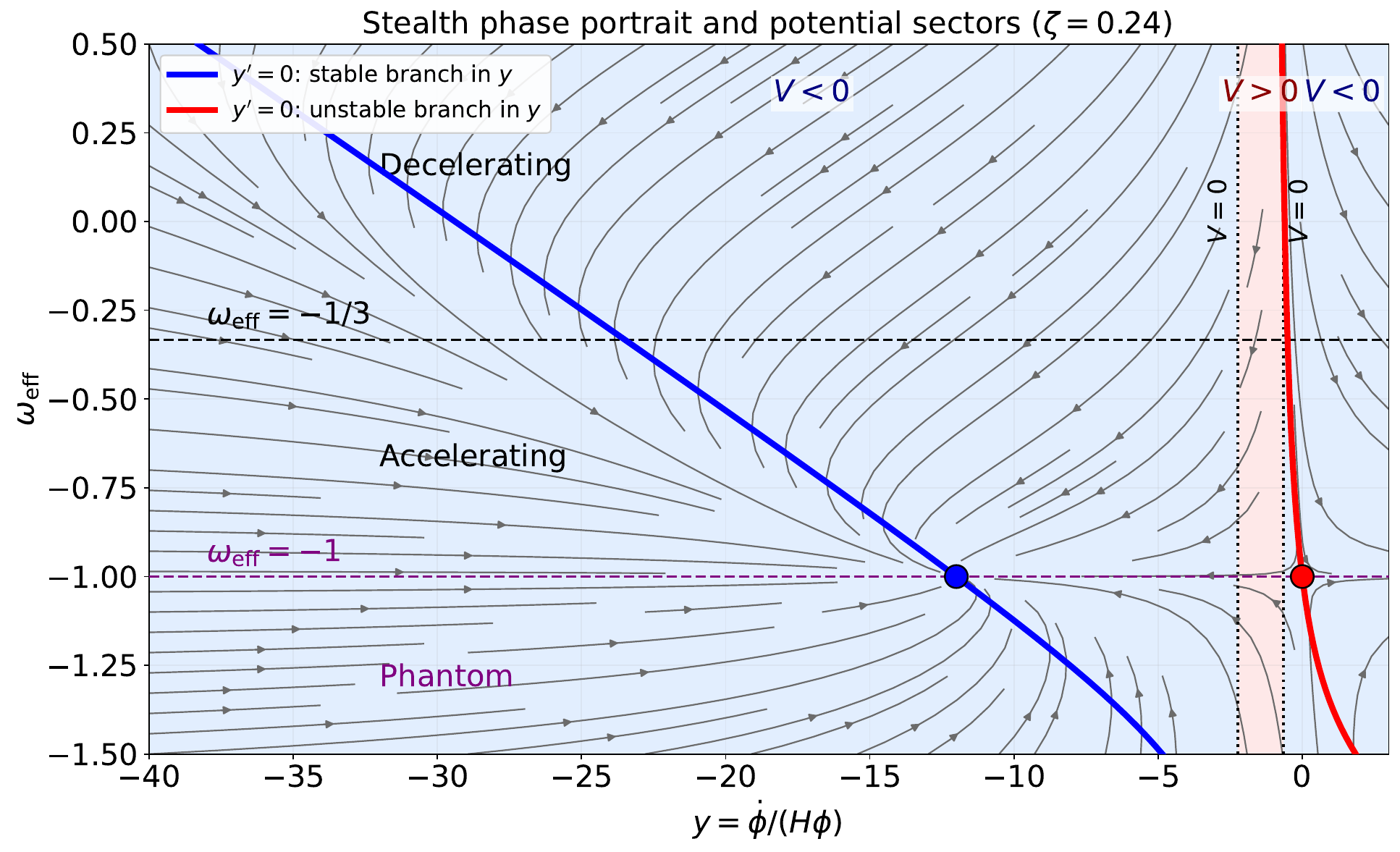}
    \includegraphics[scale=0.4]{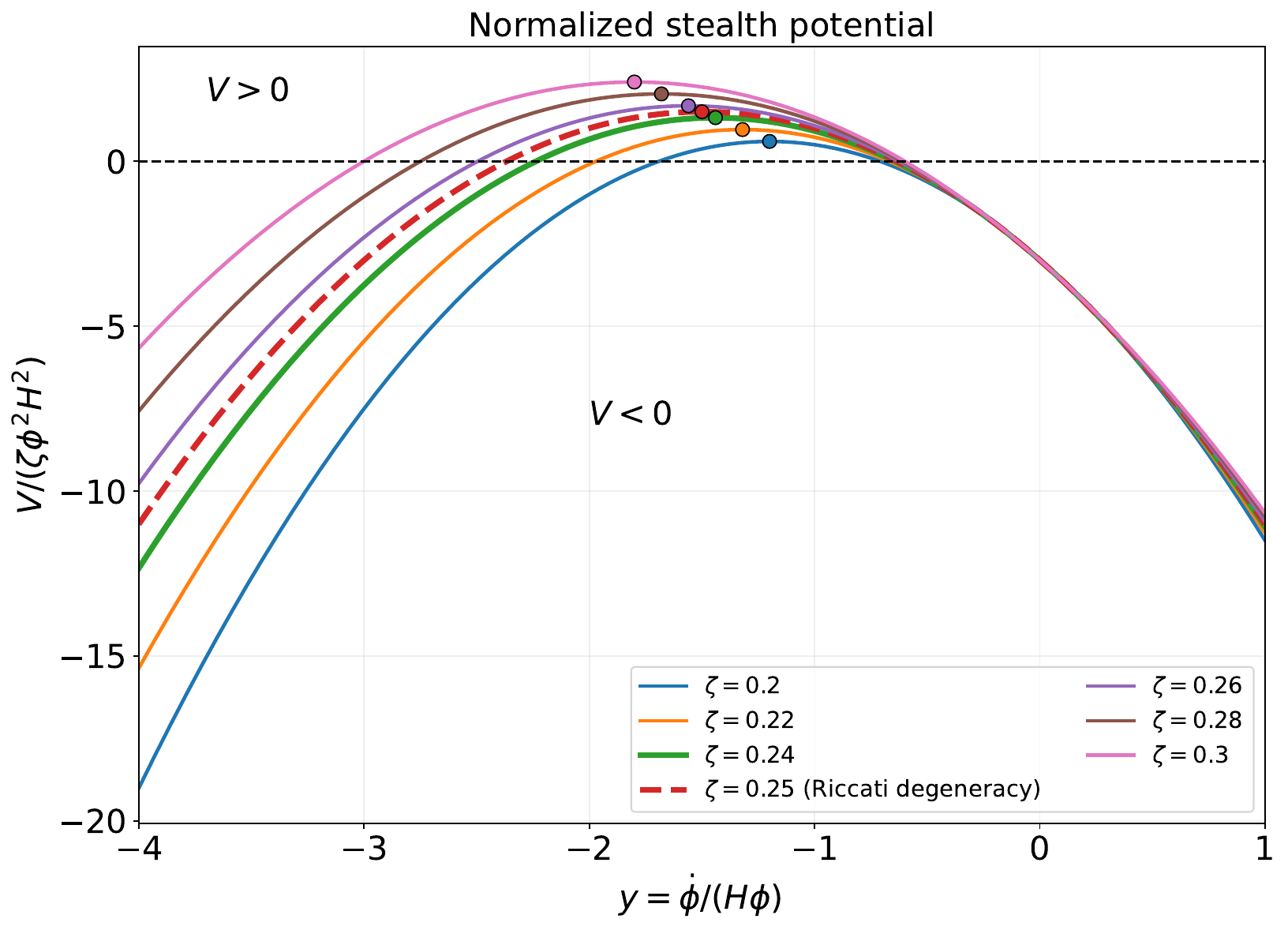}
    \caption{The upper panel represents the phase portrait of the stealth dynamical system in the $(y,\omega_{\rm eff})$ plane for $\zeta=0.24$. The color map represents the normalized potential $V/(\zeta\phi^2H^2)$. The bottom panel shows the dimensionless stealth potential as a function of $y=\dot{\phi}/(H\phi)$ for different values of $\zeta$. As $\zeta\rightarrow1/4$, the system approaches the near-critical regime. Both panels were obtained with $\alpha=1$.}
    \label{fig:phase_portraitV}
\end{figure*}
\end{center}

The normalized potential, 
\begin{equation}
    \frac{V}{\zeta\phi^2H^2}
    =
    -\left(
    6y+\frac{1}{2\zeta}y^2+3
    \right),
\end{equation}
superimposed on the phase plane in the upper panel of Fig.~\ref{fig:phase_portraitV}, shows that the displayed trajectories evolve predominantly within the \(V<0\) sector, while their passage through the \(V>0\) region is restricted to a comparatively narrow portion of the phase space. The sign of the reconstructed potential characterizes the scalar configurations required to preserve the stealth condition along the dissipative background, but it should not be interpreted as a direct measure of entropy absorption or energy transfer. As \(\zeta\to1/4\), the coefficient \(\gamma\) vanishes and the Riccati equation becomes linear. This degeneracy belongs to the scalar-field dynamics and does not correspond to a singularity or critical transition of the reconstructed potential itself, which remains a regular function of \(\zeta\) at \(\zeta=1/4\). The curve with \(\zeta=0.25\) therefore displays the potential associated with the Riccati-degeneracy case, while the neighboring curves vary smoothly with the coupling.
\onecolumngrid
\twocolumngrid

\section{Bulk Viscous Realization of the Stealth Thermodynamic Framework}
\label{sec:bulk}

The formalism developed previously remains completely general with respect to the microscopic origin of the dissipative pressure $\Pi$. However, to provide a physically meaningful cosmological realization, it is necessary to specify an explicit transport mechanism capable of generating entropy production during cosmological expansion. A natural possibility is bulk viscous cosmology, where the effective dissipative pressure is generated by departures from local thermodynamic equilibrium.  

\subsection{Constant effective equation of state}

To investigate the range of applicability of our formalism, we first examine the simplest case, namely the scenario in which $\omega_{\text{eff}}$ remains constant and is equal to $\omega _{0}$ \cite{Brevik_2011}. Eq. (\ref{eq:h_dot_general}) allows a straightforward integration, we obtain
\begin{equation}
    H(t) = \frac{1}{\frac{3}{2}(1+\omega_{0})(t-t_{0})+H^{-1}_{0}}, \label{eq:Hconst}
\end{equation}
where $H_{0}=H(t=t_{0})$. In order to simplify our notation, we define the time coordinate as $\tau = (t-t_{0})+1/(\kappa H_{0})$ with $\kappa \equiv (3/2)(1+\omega_{0})$ being a positive constant. Under this change of variable, we have $d\tau = dt$. The Riccati equation (\ref{eq:generalized_riccati}) takes the form
\begin{equation}
     \frac{dx}{d\tau} = -\gamma x^{2}+\frac{x}{\kappa \tau}+\frac{1}{\kappa\tau^{2}},\label{eq:riccatitau}
\end{equation}
where we have used our result $H(\tau) = (\kappa \tau)^{-1}$ together with Eq. (\ref{eq:effective_pressure_term}). To address Eq. (\ref{eq:riccatitau}), it is useful, as a first step, to linearize it using the standard substitution $x(\tau)=u'(\tau)/(\gamma u(\tau))$, where the prime denotes the derivative with respect to $\tau$; yielding
\begin{equation}
    \tau^{2}u''-\frac{1}{\kappa}\tau u'-\frac{\gamma}{\kappa}u=0,\label{eq:eucau}
\end{equation}
which corresponds to the Euler-Cauchy differential equation. If we propose a power law Ansatz $u(\tau) = \tau^{m}$, we obtain the following characteristic polynomial from Eq. (\ref{eq:eucau})
\begin{equation}
    m^{2}-\left(1+\frac{1}{\kappa} \right)m-\frac{\gamma}{\kappa}=0,
\end{equation}
whose solutions are given by
\begin{eqnarray}
    m_{\pm} &=& \frac{\left(\frac{5}{2}+\frac{3}{2}\omega_{0}\right)\pm \sqrt{\left(\frac{5}{2}+\frac{3}{2}\omega_{0}\right)^{2}+6\gamma \left(1+\omega_{0} \right)}}{2\kappa}, \nonumber \\ &=& \frac{\left(\frac{5}{2}+\frac{3}{2}\omega_{0}\right)\pm \sqrt{\Delta}}{2\kappa}.
\end{eqnarray}
Observe that the numerator of this latter expression is similar to the one in (\ref{eq:attractor_roots}), provided we set $\omega_{\text{eff}}=\omega_{0}$. The solution for (\ref{eq:eucau}) is simply the linear combination $u(\tau) = C_{1}\tau^{m_{+}}+C_{2}\tau^{m_{-}}$, where $C_{1,2}$ are integration constants. Going back to the variable $x$ makes it possible to write
\begin{eqnarray}
    x(\tau) &=& \frac{1}{\gamma \tau}\frac{C_{1}m_{+}\tau^{m_{+}}+C_{2}m_{-}\tau^{m_{-}}}{C_{1}\tau^{m_{+}}+C_{2}\tau^{m_{-}}},\nonumber \\ &=& \frac{\kappa}{\gamma}H(\tau)\left[\frac{C_{1}m_{+}\tau^{m_{+}}+C_{2}m_{-}\tau^{m_{-}}}{C_{1}\tau^{m_{+}}+C_{2}\tau^{m_{-}}} \right].
\end{eqnarray}
After a lengthy calculation, we can express our solution in terms of the original variable $t$, as follows
\begin{widetext}
\begin{equation}
    x(t) = H(t)\left[\frac{\frac{5}{2}+\frac{3}{2}\omega_{0}}{2\gamma}+\frac{\sqrt{\Delta}}{2\gamma}\tanh \left\lbrace \frac{\sqrt{\Delta}}{3(1+\omega_{0})}\ln \left(t-t_{0}+\frac{1}{(3/2)(1+\omega_{0})H_{0}} \right)+C_{0}\right\rbrace \right],
\end{equation}
\end{widetext}
which remains valid for $\zeta \neq 1/4$ and the constants $C_{1,2}$ have been absorbed into $C_{0}$. Since the auxiliary variable was defined as $x\equiv \dot{\phi}/\phi$, the scalar field can now be reconstructed directly from this relation.
Therefore,
\begin{equation}
    \phi(t)=\phi_0\exp\left[\int x(t)\,dt\right],
\end{equation}
where $\phi_0$ is an integration constant. Using the explicit solution for $x(t)$ and the expression obtained for $H(t)$ in \eqref{eq:Hconst}, this integral can be written in closed form. Introducing
\begin{equation}
    \tau(t)=t-t_0+\frac{2}{3(1+\omega_0)H_0},
\end{equation}
one obtains, for $\gamma\neq0$,
\begin{equation}
    \phi(\tau)
    =
    \phi_0\,
    \tau^{\frac{1}{3\gamma(1+\omega_0)}
    \left(
    \frac{5}{2}+\frac{3}{2}\omega_0
    \right)}
    \left[
    \cosh
    \left(
    \frac{\sqrt{\Delta}}{3(1+\omega_0)}
    \ln \tau
    +
    C_0
    \right)
    \right]^{\frac{1}{\gamma}},
\end{equation}
being $\phi_{0}$ the value of the scalar field at $t=t_{0}$. This expression reconstructs the stealth scalar profile compatible with the dissipative background evolution. The critical case $\gamma=0$, corresponding to $\zeta=1/4$, must be treated separately.

\subsection{Eckart model: linear dependence on the Hubble parameter}
In the simplest Eckart-like description, the bulk viscous pressure is given by
\begin{equation}
\Pi = -3\xi H,
\end{equation}
where $\xi$ denotes the bulk viscosity coefficient, which in general situations could be a function of the energy density or the Hubble parameter. Since the universe expands with $H>0$, according to Eq. (\ref{eq:entropy}) the second law of thermodynamics is automatically satisfied for $\Pi<0$ together with $\xi > 0$. From the conservation equation (\ref{eq:noncons}), we can define the total effective pressure as
\begin{equation}
P_{\rm eff}=p+\Pi,
\end{equation}
and therefore, the effective equation of state becomes
\begin{equation}
\omega_{\rm eff}=\omega - \frac{\xi}{H},
\end{equation}
where we have considered a barotropic fluid $p=\omega\rho$ and the first Friedmann equation. The cosmological evolution becomes
\begin{equation}
2\dot H +3(1+\omega)H^2 = 3\xi H.
\end{equation}
This equation explicitly shows that dissipative effects can effectively drive accelerated expansion. The condition for accelerated expansion, $\omega_{\rm eff}<-\frac13$, implies
\begin{equation}
\frac{\xi}{H}>\frac{1}{3}+\omega,
\end{equation}
Similarly, an effective phantom regime emerges whenever $\omega_{\rm eff}<-1$, which requires
\begin{equation}
\frac{\xi}{H}>1+\omega.
\end{equation}
Within this background description, the effective phantom behavior is generated by the bulk-viscous pressure rather than by assigning a wrong-sign kinetic term to the stealth scalar field. This observation concerns only the background origin of \(\omega_{\mathrm{eff}}<-1\); the stability and causal viability of the full dissipative system require a separate perturbative analysis. The Riccati equation governing the stealth dynamics is given by \eqref{eq:generalized_riccati}. Substituting the viscous pressure,
\begin{equation}
\Pi=-3\xi H,
\end{equation}
yields
\begin{equation}
\dot x + \gamma x^2-Hx-\frac32 H^2-\frac12 p = -\frac32\xi H.
\end{equation}
The dissipative contribution, therefore, enters as a negative source term in the stealth dynamics. Introducing the dimensionless variable $y\equiv x/H$, the autonomous equation becomes \eqref{eq:phase_y}, where $\gamma$ is the effective coupling constant, as before. Since $\omega_{\rm eff}=\omega-\xi/H$, the bulk viscosity coefficient directly controls the phase-space structure of the stealth sector. A particularly simple realization corresponds to a constant bulk viscosity coefficient,
\begin{equation}
\xi=\xi_0.
\end{equation}
In this case, the background evolution equation becomes
\begin{equation}
\dot H = \frac32\xi_0 H
-\frac32(1+\omega)H^2.
\end{equation}
This equation has the form of a logistic evolution equation and admits the exact solution
\begin{equation}
H(t)=
\frac{\xi_0/(1+\omega)}
{1+Ce^{-3\xi_0 t/2}},
\end{equation}
where $C$ is an integration constant fixed by the initial conditions. Consequently, the Hubble parameter asymptotically approaches the de Sitter attractor
\begin{equation}
H_{\rm dS}=\frac{\xi_0}{1+\omega}.
\end{equation}
Therefore, $\omega_{\rm eff}\to -1$ at late times. According to our previous results, the de Sitter evolution determines the stealth scalar field evolution as follows
\begin{equation}
    \dot x + \gamma x^2-H_{\rm dS}x = 0 \Rightarrow y' = -\gamma y^2+y.
\end{equation}
The critical points obtained from the condition $y'=0$ are given by
\begin{equation}
   y_{*} = 0 \qquad \text{and} \qquad y_{*} = \frac{1}{\gamma}.\label{eq:74}
\end{equation}
The nonzero root $y_{*}=1/\gamma$ is the stable attractor and is not valid at $\zeta =1/4$, which must be treated separately. In this situation, the scalar field is given by
\begin{equation}
    \phi(t) = \phi_{0}\exp \left[\frac{1}{\gamma}H_{\mathrm{dS}}(t-t_{0}) \right].
\end{equation}
Observe that, at this point, the stealth scalar field is required to evolve exponentially in time, in step with the scale factor evolution, $a(t) \propto \exp(H_{\mathrm{dS}}t)$, so that it functions as a thermodynamic tracker and continues to satisfy its stealth condition. Then  the stealth field evolves consistently with the entropy producing background. Because the selected attractor has \(\dot{\phi}\neq0\), it corresponds to the dynamically evolving stealth branch, whereas the static solution \(y_*=0\) is a repeller of the background thermodynamic dynamics. The perturbative stability of either branch cannot be inferred from this background analysis alone and requires an examination of the complete coupled perturbation system.

\section{Scalar perturbations of the dissipative stealth configuration}
\label{sec:pert}
In the Newtonian gauge, the FLRW metric is perturbed by two scalars $\Phi$ and $\Psi$. Then we have the following for the line element
\begin{equation}
    ds^{2}=-(1+2\Phi)dt{2}+a^{2}(t)(1-2\Psi)\delta_{ij}dx^{i}dx^{j},
\end{equation}
we consider perturbations around a homogeneous background scalar field,
\begin{equation}
\phi(t,\mathbf{x})=\phi_0(t)+\delta\phi(t,\mathbf{x}),
\end{equation}
and decompose the scalar perturbation into Fourier modes, \(\delta\phi\propto\exp(i\mathbf{k}\cdot\mathbf{x})\), where \(\mathbf{k}\) is the comoving wave vector and \(k=\lvert\mathbf{k}\rvert\) is its magnitude, a related analysis can be found in Ref. \cite{Blanco2026}. For a non-minimally coupled scalar field, we have
\begin{equation}
    \Box \phi = \zeta R\phi+V'(\phi),\label{eq:kg}
\end{equation}
with $R$ being the Ricci scalar, and the prime denoting the derivative with respect to the scalar field. The $\Box$ denotes the d'Alembertian operator, which is defined as $\Box \phi =(1/\sqrt{-g})\partial_{\mu}(\sqrt{-g}g^{\mu \nu}\partial_{\nu}\phi)$, where $g$ is the determinant of the metric. The first-order perturbation of Eq. (\ref{eq:kg}),
$$
\delta(\Box\phi)=\delta\!\left[\zeta R\phi+V'(\phi)\right],
$$
gives
\begin{equation}
    \delta(\Box \phi) = -\delta \ddot{\phi} - 3H\delta \dot{\phi}-\frac{k^{2}}{a^{2}}\delta \phi + \underbrace{\dot{\phi}_{0}(\dot{\Phi}+3\dot{\Psi})+2\Phi(\ddot{\phi}_{0}+3H\dot{\phi}_{0})}_{\mathcal{T}_{m}},
\end{equation}
The term \(k^2\delta\phi/a^2\) arises from the spatial derivatives of the scalar perturbation. The first-order perturbation of the right-hand side of Eq. (\ref{eq:kg}) is
\begin{equation}
    \delta (\zeta R\phi+V'(\phi)) = \zeta \phi_{0}\delta R + \left[\zeta R_{0}+V''(\phi_{0}) \right]\delta \phi,  
\end{equation}
where we have considered fluctuations of the Ricci scalar due to its dependence on the metric. The background equation allows us to write the second derivative of the potential as follows
\begin{eqnarray}
    V''(\phi_{0}) &=& \frac{dV'(\phi_{0})}{d\phi_{0}}  = \frac{1}{\dot{\phi_{0}}}\frac{d}{dt}V'(\phi_{0})\nonumber \\ &=& -\frac{1}{\dot{\phi_{0}}}\frac{d}{dt}\lbrace \ddot{\phi_{0}}+3H\phi_{0}+\zeta R_{0}\phi_{0}\rbrace,
\end{eqnarray}
then the curvature of the potential is determined by the background quantities. Therefore, we can write
\begin{equation}
    \delta \ddot{\phi} +3H\delta\dot{\phi}+\left(\frac{k^{2}}{a^{2}}+M^{2}\right)\delta \phi = \mathcal{T}_{m}-\zeta \phi_{0}\delta R,
\end{equation}
where we have defined $M^{2}\equiv \zeta R_{0}+V''(\phi_{0})$. The Einstein equations in this setup can be expressed as $G_{\mu \nu}=T_{\mu \nu}+T^{(S)}_{\mu \nu}$, where $T^{(S)}_{\mu \nu}$ denotes the energy-momentum tensor of the stealth configuration and $T_{\mu \nu}$ corresponds to the matter sector. Taking the trace of this relation yields $-R=T+T^{(S)}$. For a dissipative fluid, we have $T=-\rho+3P$, where $P\equiv p+\Pi$ and $p$ being the equilibrium pressure. The background stealth condition leads to $T^{(S)}=0$, yielding
\begin{equation}
    R_{0} = \rho_{0}-3(p_{0}+\Pi_{0}).
\end{equation}
At the background level, the stealth field does not affect the scalar curvature. However, its perturbation is nonzero ($T^{(S)}_{\mu \nu} \propto \zeta \delta G_{\mu \nu}$, for $\zeta=0$ this condition is lost), so
\begin{equation}
    \delta R = \delta \rho -3\delta p-3\delta \Pi-\delta T^{(S)},
\end{equation}
the backreaction of the field ($\delta T^{(S)}$) depends on $\delta \phi$, $\Phi$, and $\Psi$. Additionally, the dissipative pressure contributes to the perturbation of the scalar curvature. We can write
\begin{align}
     & \delta \ddot{\phi} +3H\delta\dot{\phi}+\left(\frac{k^{2}}{a^{2}}+M^{2}\right)\delta \phi = \nonumber \\ 
     & \mathcal{T}_{m}-\zeta \phi_{0}\left[\delta \rho -3\delta p-3\delta \Pi\right] + \zeta \phi_{0}\delta T^{(S)}.
     \label{eq:full}
\end{align}
As can be seen, the scalar perturbation couples directly to the dissipative sector. The perturbed Klein-Gordon equation is sourced by the term $3\zeta \phi_{0}\delta \Pi$. This coupling is present only for a non-minimal coupling ($\zeta \neq 0$). Since the entropy production rate obeys $T\dot{S}=-3HV\Pi$, its perturbation is proportional to $\delta \Pi$; the stealth field is thus coupled, at first order, to the entropy production. To obtain an explicit evolution equation for \(\delta\Pi\), we adopt the truncated Israel-Stewart transport equation. More general causal dissipative models may be considered, but they lie beyond the scope of the present analysis. In the truncated formulation, the bulk-viscous pressure satisfies \cite{maartens}
\begin{equation}
    \tau u^{\mu}\nabla_{\mu}\Pi+\Pi=-\xi \Theta, \quad \Theta=\nabla_{\mu}u^{\mu}, \label{eq:is}
\end{equation}
where \(\tau\) is the relaxation time, \(\xi\) is the bulk-viscosity coefficient, and \(u^\mu\) is the fluid four-velocity satisfying \(u^\mu u_\mu=-1\). On the homogeneous FLRW background, \(\Theta_0=3H\), and the transport equation reduces to
\begin{equation}
    \tau_{0}\dot{\Pi}_{0}+\Pi_{0} = -3\xi_{0}H,
\end{equation}
In general, both \(\tau\) and \(\xi\) may depend on the fluid energy density. The Eckart constitutive relation is formally recovered in the limit \(\tau\to0\).
We now perturb the fluid variables. The four-velocity can be written as \(u^\mu=(u^0,u^i)\), where \(u^i\) is a first-order peculiar velocity. The normalization condition \(g_{\mu\nu}u^\mu u^\nu=-1\) gives \(u^0\simeq1-\Phi\). Using
$$
\Theta=\nabla_\mu u^\mu
=\frac{1}{\sqrt{-g}}\partial_\mu\!\left(\sqrt{-g}\,u^\mu\right),
$$
together with \(\sqrt{-g}\simeq a^3(1+\Phi-3\Psi)\), the temporal contribution becomes
$$
\partial_0\!\left(\sqrt{-g}\,u^0\right)
\simeq
3a^3H-9a^3H\Psi-3a^3\dot{\Psi}.
$$
For the spatial contribution, we define \(\theta\equiv\partial_i u^i\), so that
$$
\partial_i\!\left(\sqrt{-g}\,u^i\right)\simeq a^3\theta.
$$
Combining both contributions yields
\begin{align}
    & \Theta \simeq a^{-3}(1-\Phi+3\Psi)a^{3}(3H-9H\Psi-3\dot{\Psi}+\theta),\nonumber \\ 
    & \simeq 3H+\theta -3H\Phi - 3\dot{\Psi}.
\end{align}
Since the background expansion scalar is \(\Theta_0=3H\), its first-order perturbation is
\begin{equation}
    \delta \Theta = \theta -3H\Phi - 3\dot{\Psi},
\end{equation}
here, \(\theta=\partial_i u^i\) denotes the scalar velocity-divergence perturbation. We now perturb Eq. (84), allowing both transport coefficients to depend on the energy density, \(\tau=\tau(\rho)\) and \(\xi=\xi(\rho)\). Consequently, \(\delta\tau=\tau_{\rho}\delta\rho\) and \(\delta\xi=\xi_{\rho}\delta\rho\), where the subscript \(\rho\) denotes differentiation with respect to the energy density.

We have $\delta \tau = \tau_{\rho}\delta \rho$ and $\delta \xi = \xi_{\rho}\delta \rho$, where the subscript denotes the derivative with respect to $\rho$, yielding
\begin{equation}
u^{\mu}\nabla_{\mu}\Pi=u^{0}\partial_{0}\Pi+u^{i}\partial_{i}\Pi.
\end{equation}
Decomposing the bulk-viscous pressure as \(\Pi=\Pi_0(t)+\delta\Pi\), we have
$$
\partial_0\Pi=\dot{\Pi}_0+\delta\dot{\Pi},
\qquad
\partial_i\Pi=\partial_i\delta\Pi.
$$
Because both \(u^i\) and \(\partial_i\delta\Pi\) are first-order quantities, their product contributes only at second order and must be neglected. Therefore,
\begin{equation}
u^\mu\nabla_\mu\Pi
\simeq
\dot{\Pi}_0+\delta\dot{\Pi}-\Phi\dot{\Pi}_0.
\label{eq:perturbed_convective_derivative}
\end{equation}
Using $\tau = \tau_{0}+\delta \tau$, we can write
\begin{align}
    & \tau u^{\mu}\nabla_{\mu}\Pi + \Pi = (\tau_{0}+\delta \tau)(\dot{\Pi}_{0} +\delta \dot{\Pi} -\Phi \dot{\Pi}_{0}) + \Pi_{0}+\delta \Pi, \nonumber \\
    & \simeq \tau_{0}\dot{\Pi}_{0} +\Pi_{0}+ \tau_{0}\delta \dot{\Pi}-\tau_{0}\Phi \dot{\Pi}_{0}+\tau_{\rho}\dot{\Pi}_{0}\delta \rho + \delta \Pi,
\end{align}
where the first two terms on the right hand side correspond to the background case. Finally, we obtain 
\begin{equation}
    \tau_{0}\dot{\Pi}_{0} +\Pi_{0}+\tau_{0}\delta \dot{\Pi}-\tau_{0}\Phi \dot{\Pi}_{0}+\tau_{\rho}\dot{\Pi}_{0}\delta \rho + \delta \Pi = -(\xi_{0}+\delta \xi)(\Theta_{0}+\delta \Theta)
\end{equation}
which, at first order, reads
\begin{equation}
\tau_0 \delta\dot{\Pi}+\delta\Pi
=
-\xi_0\delta\Theta
-\left(\xi_\rho\Theta_0+\tau_\rho\dot{\Pi}_0\right)\delta\rho
+\tau_0\Phi\dot{\Pi}_0 .
\label{eq:perturbed_IS}
\end{equation}
A few remarks are needed to close this section. The presence of $3\zeta \phi_{0}\delta \Pi$ can breaks the background degeneracy. Even if different models share the same background expansion history \(H(t)\), their perturbative dynamics need not be identical. For a perfect fluid, \(\delta\Pi=0\), whereas in a dissipative fluid the term \(3\zeta\phi_0\delta\Pi\) provides an additional source for the stealth perturbation. Therefore, the dissipative and non-dissipative scenarios can, in principle, be distinguished at the perturbative level even when they reproduce the same background evolution. Determining the resulting modification of matter growth requires solving the complete coupled system of scalar, metric, matter, and dissipative perturbations.

\section{Conclusions}
\label{sec:final}
In this work, we have established a novel connection between the cosmological stealth scalar field and non-equilibrium thermodynamics. To our knowledge, the description of stealth configurations within the framework of irreversible thermodynamic processes has not been previously explored in the literature. By considering a non-minimally coupled scalar field evolving on a dissipative FLRW background, we demonstrated that the stealth condition manifests as a generalized Riccati equation governing the kinematical evolution of the field. In this framework, the dissipative pressure of the cosmic fluid serves as a key thermodynamic quantity. As a result, the scalar field acts as a tracer of the system’s irreversible evolution. 

A main result of our analysis is the global structure description of the system, mapped through a two-dimensional phase space formulation in the $(y,\omega_{\mathrm{eff}})$ plane. As found, the thermodynamic state of the model admits two branches, crucially, the stability analysis proves that the irreversible arrow of time dynamically selects the stable attractor branch and remains consistent with the thermal evolution of the universe. From a phenomenological perspective, this thermodynamic construction offers significant advantages over the standard scalar-tensor approaches. The framework allows for an effective phantom regime driven entirely by macroscopic dissipative processes, without introducing a fundamental scalar degree of freedom with an explicitly wrong-sign kinetic term. This statement concerns the origin of the phantom background evolution and should not be interpreted as proof of perturbative stability. A relevant feature of our formulation is that the stealth potential is not imposed a priori; rather, it is reconstructed dynamically so that the scalar field remains gravitationally invisible throughout the background evolution. The resulting potential landscape depends on the cosmological evolution and on the non-minimal coupling parameter.
As illustrated by the phase-space structure, the displayed trajectories evolve predominantly within the negative-potential sector, reflecting the form of the potential required to preserve the stealth configuration along the dissipative background evolution.
For the trajectories displayed in Fig. (\ref{fig:phase_portraitV}), the region with positive potential occupies only a limited portion of the explored phase space. As \(\zeta\) approaches \(1/4\), the scalar-field dynamics become increasingly asymmetric because one of the Riccati branches moves toward large \(\lvert y\rvert\), while the reconstructed potential itself remains regular at this coupling.

A further central result concerns the cosmological perturbations. Although the stealth field is invisible on the background, this invisibility does not generally persist at first order. The scalar perturbation couples directly to the dissipative-pressure perturbation through a term that is present only for non-minimal coupling. Since the dissipative pressure obeys its own evolution equation, it introduces an additional perturbative degree of freedom. Therefore, the dissipative stealth can, in principle, be distinguished from a non-dissipative source with the same background expansion history through its perturbative dynamics. In this sense, the physically distinctive content of the model resides in its perturbation sector, whereas the irreversible background evolution selects the dynamically stable stealth branch. Whether this selected branch is also free of ghost and gradient instabilities remains to be determined from the complete coupled perturbation system.

Finally, the consistency and applicability of the background formalism are illustrated through two explicit bulk-viscous realizations: a constant effective equation-of-state scenario and an Eckart-type dissipative model. In both cases, the stealth scalar profile can be exactly reconstructed as a function of time. Furthermore, as entropy production dynamically drives the cosmic expansion toward an asymptotic de Sitter state, the scalar field naturally adapts, evolving in parallel with the background to sustain its stealth configuration. While we illustrated our formalism using constant viscosity and a first-order Eckart model, future work will address more general non-equilibrium scenarios. Specifically, extending stealth kinematics to causal thermodynamic frameworks, such as the Israel-Stewart formalism, see \cite{CRUZ2017159, Cruz_2017} and references therein, is necessary to overcome some issues inherent to first-order theories. Along the same line, the perturbative viability of the dissipative stealth, in particular the absence of ghost or gradient instabilities, could be examined by recasting the model as a scalar-tensor theory following the stealth stability analysis of \cite{faraoni}. Furthermore, a dynamic bulk viscosity coefficient dependent on energy density or the Hubble parameter could reveal richer phase-space topologies. The implementation of these extended thermodynamic models remains an open avenue for future research.

\section*{Acknowledgments}
The work of M. C., C. C. V. and G. A. P. received partial support from S.N.I.I. (SECIHTI-M\'exico), and they also express their gratitude for the encouragement provided by ProDeP-M\'exico, CA-UV-320: \'Algebra, Geometr\'\i a y Gravitaci\'on. G.A.P. was supported by SECIHTI through the {\it Estancias Posdoctorales por México 2023(1)} program. J. S. acknowledges the FONDECYT grant N°1220065, Chile. V. H. C. acknowledges CEFITEV-UV for partial support.
\bibliographystyle{apsrev4-1}
\bibliography{biblio}
\end{document}